\documentclass[aps,prl,twocolumn,floatfix,footinbib,showpacs,showkeys,superscriptaddress]{revtex4-1} %I
\usepackage[english]{babel}
\usepackage[utf8]{inputenc}
\usepackage{graphicx}
\usepackage{fancyhdr}
\usepackage[active]{srcltx}
\usepackage{setspace}
\usepackage{float}
\usepackage{amsmath}
\usepackage{amsfonts}
\usepackage{natbib}%
\usepackage{color}
\usepackage{bm} %I
\begin{document}

\title{Wurtzite spin lasers}

\author{Paulo~E. Faria~Junior}
\affiliation{S\~ao Carlos Institute of Physics, University of S\~ao Paulo, 13566-590 S\~ao Carlos, S\~ao Paulo, Brazil}
\author{Gaofeng Xu}
\affiliation{Department of Physics, University at Buffalo, State University of New York, Buffalo, New York 14260, USA}
\author{Yang-Fang Chen}
\affiliation{Department of Physics, National Taiwan University, Taipei 106, Taiwan}
\author{Guilherme~M. Sipahi}
\affiliation{S\~ao Carlos Institute of Physics, University of S\~ao Paulo, 13566-590 S\~ao  Carlos, S\~ao Paulo, Brazil}
\affiliation{Department of Physics, University at Buffalo, State University of New York, Buffalo, New York 14260, USA}
\author{Igor \v{Z}uti\'{c}}
\email{zigor@buffalo.edu}
\affiliation{Department of Physics, University at Buffalo, State University of New York, Buffalo, New York 14260, USA}

%===============================================================================
   
\begin{abstract}
Semiconductor lasers are strongly altered by adding spin-polarized carriers. Such spin lasers  could overcome 
many limitations of their conventional (spin-unpolarized) counterparts. While the vast majority of experiments 
in spin lasers employed zinc-blende semiconductors, the room temperature electrical manipulation was first 
emonstrated in wurtzite GaN-based lasers. However, the underlying theoretical description of wurtzite spin 
lasers is still missing. To address this situation, focusing on (In,Ga)N-based wurtzite quantum wells, 
 we develop a theoretical framework in which the calculated microscopic spin-dependent gain is combined 
 with a simple rate equation model. A small spin-orbit coupling in these wurtzites supports simultaneous 
 spin polarizations of electrons and holes, providing unexplored opportunities to control spin lasers. 
 For example, the gain asymmetry, as one of the key figures of merit related to spin amplification, 
 can change the sign by simply increasing the carrier density. The lasing threshold reduction has 
 a nonmonotonic depenedence on electron spin polarization, even for a nonvanishing hole spin polarization. 
\end{abstract}
\maketitle

%===============================================================================
%I nitrides vs wurtzites, in various places
%I is it possible to get the correct format for new PRB so that we do not need
% to use onlinecite, but Section numbers are PRB like?????

\section{I. Introduction}
%TO DO
% Add PRL02, Hirohata, older spintronics reviews,...
%CHECK what is consistent with Koch 1999 %note APL 2008

%----------------------
% abbreviations:
% ZB: zinc-blende
% WZ: wurtzite
% SOC: spin-orbit coupling
% QW: quantum well
% VCSEL: vertical cavity surface emitting laser
% conduction band: CB
% heavy hole: HH
% light hole: LH
% split-off hole: SO
% crystal-field split-off hole: CH

% Auger is negligible in GaN-based QWs
%\bibitem{Hader2008:APL}
%J.~Hader, J.~V.~Moloney, B.~Pasenow, S.~W.~Koch, M.~Sabathil, N.~Linder, and S.~Lutgen,
%Appl. Phys. Lett. {\bf 92}, 261103 (2016).

Introducing spin-polarized carriers in semiconductor lasers offer an alternative
path to realize spintronic applications, beyond the usually employed magnetoresistive
effects~\cite{Hallstein1997:PRB,Ando1998:APL,Rudolph2003:APL, Holub2007:PRL,Hovel2008:APL,Basu2008:APL,%
Basu2009:PRL,Saha2010:PRB,Fujino2009:APL, Jahme2010:APL,Gerhardt2011:APL,Iba2011:APL,Frougier2013:APL,%
Frougier2015:OE,Alharthi2015:APL,Alharthi2015:APLb,Hsu2015:PRB}.
%Through conservation of angular moment, the angular moment
%of the spin-polarized carriers is transferred to photons and thus, through carrier recombination, leads
%to the circularly polarized emitted light~\cite{Sinova2012:NM}.
Through carrier recombination, the angular momentum of the spin-polarized carriers 
is transferred to photons, thus leading to the circularly polarized emitted light~\cite{Sinova2012:NM}. %FJ10
Such spin lasers provide opportunity to extend the functionality of spintronic devices,
as well to exceedthe performance of conventional (spin-unpolarized) lasers, from reducing the 
lasing threshold to improving their dynamical performance and digital 
operation~\cite{Lee2010:APL,Boeris2012:APL,Lee2012:PRB,Lee2014:APL,Wasner2015:APL,FariaJunior2015:PRB,Hopfner2014:APL,%
Lindemann2016:APL,Pusch2015:EL}.

Almost all spin lasers have been based on  zinc-blende (ZB) semiconductors, such
as GaAs or InAs, in which spin-dependent optical transitions were extensively studied
for over 45 years~\cite{Meier:1984}. However, a lone exception of a spin laser with an gain (active)
region made of a wurtzite (WZ) semiconductor has so far also been the only case
of an electrically manipulated %I9 operated 
spin laser at room temperature~\cite{Chen2014:NN,Zutic2014:NN}. 
Unlike many theoretical studies of ZB spin lasers~\cite{Gothgen2008:APL,Holub2011:PRB,FariaJunior2015:PRB}, 
a theoretical description for WZ spin lasers is still missing. Focusing on WZ GaN-based quantum wells (QWs) 
as the gain region, we develop the first microscopic description of  WZ spin lasers.

\begin{figure}[h!]
\begin{center}
\includegraphics{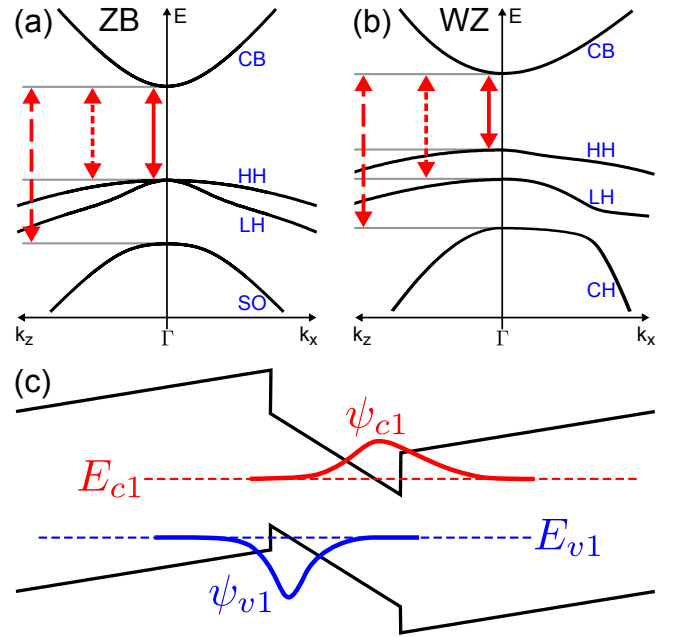}
\caption{Band structure for  (a) zinc-blende (ZB) and wurtzite (WZ) bulk semiconductor.  Conduction band (CB), and the 
valence band (VB):  light, heavy holes (HH, LH), spin-orbit spin-split off holes (SO),  and crystal-field split-off holes (CH).
Arrows: 
possible  band edge  $ {\bm k}=0$ 
optical transitions, weighted by different
coefficients, discussed in the main text.
(c) Quantum confined Stark effect due to
spontaneous and piezoelectric polarization in WZ quantum wells. First
CB, VB energy levels and the envelope functions are depicted.}
\label{fig:ZBWZpol}
\end{center}
\end{figure}

The significance of WZ materials for optical devices has been recognized by 
the 2014 Nobel prize in physics for an efficient blue light emitting diodes (LEDs).
WZ-based optical devices using a direct band gap GaN and its In and Al alloys are 
ubiquitous in our daily lives, from efficient lightning to blue-ray disc readers. 
Due to their high electron saturation velocities and high breakdown voltages,
GaN-based semiconductors are also promising for high-speed/high-power 
electronic devices~\cite{Akasaki2015:RMP, Amano2015:RMP,Nakamura2015:RMP}.
However, for spin-dependent optical properties, WZ GaN does not appear encouraging,
leading to only a negligibly small degree of a circular polarization of an emitted
light which can be attributed to a rather weak spin-orbit coupling (SOC)~\cite{Chen2005:APL}. 
Therefore, the realization of the first electrically manipulated %I9 operated 
spin laser at room temperature
using GaN-based gain region came as a surprise.

To better understand the differences between employing ZB and  WZ semiconductors in optical devices, 
in Figs.~\ref{fig:ZBWZpol}(a) and (b) we show
their bulk band structure and possible band edge optical transitions within the
conventional $8\times8$ $\bm{k{\cdot}p}$ Hamiltonians, using the typical notation:
conduction band (CB),  heavy holes (HH),  light holes (LH) and spin-orbit split-off holes (SO) for 
ZB~\cite{Winkler:2003,Lee2014:PRB} %I13
and CB, HH, LH and crystal-field split-off hole (CH) for WZ~\cite{Chuang1996:PRB}. 
Each of the  marked dipole transitions has a different amplitude for specific spins that
apply both to radiative recombinations and excitations. 

If we denote the photon density of positive (negative) helicity by $S^+$ ($S^-$),  
we can describe the relevant helicity in each of the transitions. 
For example, in  the CB-HH transition  spin up (down) leads to $S^-$ ($S^+$), in CB-LH
spin up (down) leads to $aS^+$ ($aS^-$), while in  CB-SO for ZB, or CB-CH for WZ,  
spin up (down) leads to $bS^+$ ($bS^-$). 
For ZB  the amplitude of helicity contributions are fixed: $a=1/3$, $b=2/3$. The 
electron spin polarization
in terms of spin up (down) electron density $n_+$ ($n_-$), 
\begin{equation}
P_n=(n_+-n_-)/(n_+ + n_-), 
\label{eq:Pn}
\end{equation}
arising from optical spin injection (HH/LH-CB) yields $P_n= (1-1/3)(1+1/3) = 50\%$, 
a well-known result  
at the band gap, neglecting electron spin relaxation~\cite{Pierce1975:PLA,Zutic2004:RMP}.
In contrast,  for WZ the corresponding amplitudes depend on the materials parameters related 
to the SOC~\cite{Chuang1996:PRB},
\begin{equation}
a = E^2_+/(E^2_+ + 2\Delta^2_3), \quad   b=2\Delta^2_3 / (E^2_+ + 2\Delta^2_3),
\label{eq:abWZ}
\end{equation}
where the energy $E_+$ is expressed as,
\begin{equation}
E_+ = (\Delta_1 - \Delta_2)/2  + \sqrt{(\Delta_1 - \Delta_2)^2/4 + 2\Delta^2_3} \, ,
\end{equation}
in terms of the crystal field splitting energy  $\Delta_1$ 
and SOC splitting energies $\Delta_{2,3}$. 
With removed HH and LH degeneracy at the wavevector ${\bm k=0}$ ($\Gamma$-point) in WZ semiconductors
[see Fig.~\ref{fig:ZBWZpol}(b)], one would expect  $P_n \rightarrow 100$\% optical spin injection  at the band gap, 
overcoming the 50\% limitation of ZB materials~\cite{Zutic2004:RMP}. 
However, due to  the relatively weak SOC in nitride-based materials~\cite{note:GaN}, the energy separation 
for the topmost valence bands is typically $\sim$10-20 meV, comparable to  
the energy of the broadening effects by impurities and room temperature, 
ultimately leading to inefficient optical spin injection~\cite{Chen2005:APL}. 
In GaN-based spin LEDs only a small circular polarization of electroluminescence  was detected at 200 K~\cite{Banerjee2013:APL}
as well as at 300 K in the  applied magnetic field~\cite{Chen2010:APL}. %I12
These limitations could be overcome in  electrical spin injection or extraction, %~\cite{Zutic2002:PRL}, 
as shown (In,Ga)N/GaN-based %I9
nanodiscs  and nanorods covered by  Fe$_\textrm{3}$O$_\textrm{4}$ 
nanoparticles~\cite{Chen2014:NL, Chen2014:NN,Zutic2014:NN}.

In this study, we investigate WZ spin lasers with $\textrm{In}_{0.1}\textrm{Ga}_{0.9}\textrm{N}$/GaN QWs 
as their gain region using microscopic  $\bm{k{\cdot}p}$ band structure calculations. While a weak SOC retains
desirable spin-dependent properties of optical gain, it also necessitates simultaneous consideration of electron 
and hole spin polarizations, largely overlooked in the previous studies. By combining macroscopic rate 
equations with microscopic gain calculations based on a $\bm{k{\cdot}p}$ method,  we establish a versatile 
method to describe spin lasers which extends the strengths of the two complementary approaches.  

In Sec.~II we describe the $\bm{k{\cdot}p}$ method to evaluate the electronic
structure of (In,Ga)N QW which is used in Sec.~III to calculate microscopic spin-dependent optical gain. 
In Sec.~IV we combine these microscopic gain calculations with simple rate equations, suitable to 
describe various dynamical phenomena in spin lasers. In Sec.~V we discuss future opportunities
to apply our theoretical framework to other systems.

%===============================================================================

\section{II. Quantum well electronic structure}
\label{sec:QW_bs}

An important  consequence of  the atomic arrangement of WZ materials
along the [0001] direction is the presence of the polarization fields.
A relative displacement between cations and anions in the unit cell leads to the 
spontaneous polarization along the growth direction in QWs.  
Under external 
applied strain this cation-anion displacement  is modified and 
also yields piezoelectric polarization~\cite{note:pols}. 
Such polarization fields are schematically shown in  Fig.~\ref{fig:ZBWZpol}(c) 
for a WZ QW. The response of the quantum confined states to the static electric field
is known as the quantum confined Stark effect and recognized as a very efficient 
mechanism to tune the optical transitions in semiconductor nanostructures~\cite{Chuang:2009}. %I Gywat Book
These polarization fields modify both electronic levels as well as change the  spatial 
electron-hole separation and thus the overlap integral between CB and VB wave functions.

Within the $\bm{k{\cdot}p}$ method 
combined with the envelope function approximation, and 
including the polarization effects, the total Hamiltonian of the QW system is, 
\begin{equation}
H_{\textrm{QW}}(z) = H_{kp}(z) + H_{\textrm{st}}(z) + H_{\textrm{O}}(z) + H_{\textrm{pol}}(z) \, ,
\label{eq:HQW}
\end{equation}
with the growth axis along the $z$ direction (the $c$ axis, or [0001] direction, of the
WZ structure). The Hamiltonian $H_{kp}(z)$ denotes the $\bm{k{\cdot}p}$ term, $H_{\textrm{st}}(z)$ 
describes the strain, 
$H_{\textrm{O}}(z)$ includes the band-offset at the interface 
that generates the QW energy profile,  and $H_{\textrm{pol}}(z)$ includes the potential
profile due to spontaneous and piezoelectric polarizations. In this study, we considered 
the $8\times8$ $\bm{k{\cdot}p}$ Hamiltonian for WZ materials with explicit interaction between 
CB and VB which gives rise to the dipole coupling for optical  transitions. The specific definitions of these Hamiltonians 
are described in Refs.~\cite{FariaJunior2012:JAP,Miao2012:PRL,FariaJunior2014:JAP}. In order 
to numerically solve the resulting system of coupled differential equations from Eqs.~(\ref{eq:HQW}), we apply the 
plane wave expansion discussed in Refs.~\cite{Sipahi1996:PRB,Lee2014:PRB,%I9
FariaJunior2012:JAP,FariaJunior2014:JAP,FariaJunior2015:PRB}.

\begin{figure}[h!]
\begin{center}
\includegraphics{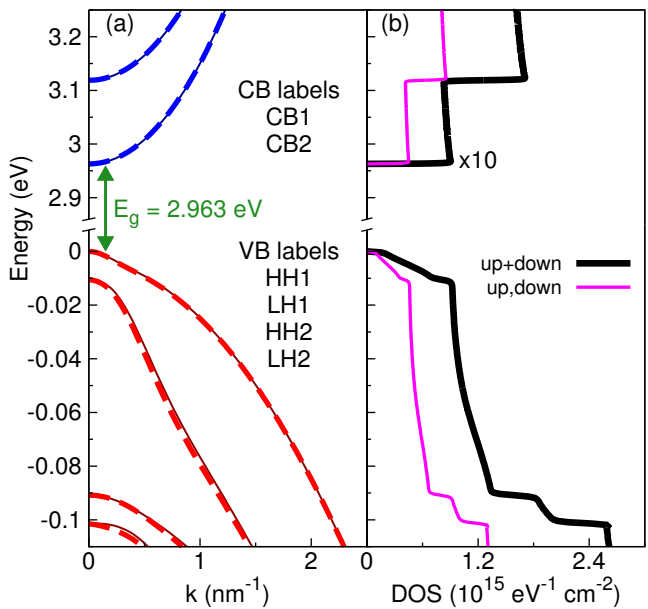}
\caption{(a) Band structure for the $\textrm{In}_{0.1}\textrm{Ga}_{0.9}\textrm{N}$/GaN
QW along an in-plane $k$-direction. Solid and dashed lines show the spin splitting
due to the asymmetric polarization potential. (b) Density of states (DOS) calculated 
from the band structure in (a). The up and down labels refer to spin. 
The CB DOS is multiplied by a factor of 10 to match the VB
scale. The band gap is $E_g=2.963\; \textrm{eV}$ (CB1-HH1 energy difference).}
\label{fig:bs_dos}
\end{center}
\end{figure}
% 2.963 eV ~ 418 nm
% 3 eV ~ 413 nm

For the gain region of the laser we consider a  3 nm thick single strained $\textrm{In}_{0.1}\textrm{Ga}_{0.9}\textrm{N}$
QW,  surrounded by 6 nm GaN barriers, the typical lengths and composition of  (In,Ga)N-based vertical cavity surface 
emitting lasers (VCSELs)~\cite{Lu2008:APL,Lu2010:APL,Kasahara2011:APE,Lin2014:LPL}.
The bulk InN and GaN parameters are obtained from Ref.~\cite{Miao2012:PRL}, we use their  linear interpolation
for the alloy $\textrm{In}_{0.1}\textrm{Ga}_{0.9}\textrm{N}$ and the bowing parameter for the 
band gap, $E_g$~\cite{Vurgaftman2001:JAP}. The interface band offsets  
are $\Delta E_C = 0.7\Delta E_g$ 
and $\Delta E_V = 0.3\Delta E_g$~\cite{Dong2013:JAP}.  We choose  $E_g$ at T = 300 K with 
Varshni parameters and the refractive indexes from Refs.~\cite{Vurgaftman2001:JAP, note:web}.

To develop some intuition about the relevant SOC parameters in (In,Ga)N QWs,  
we recall that in GaAs, as the representative ZB semiconductor, at the
$\Gamma$-point HH and LH are degenerate and separated by 
$\Delta_{SO} = 0.341$ eV~\cite{Vurgaftman2001:JAP} from the SO band. % FJ11
It is helpful to think of ZB GaAs as a WZ structure without 
crystal-field splitting energy ($\Delta_1 = 0$) and a much larger SOC that yields 
$\Delta_2 = \Delta_3 = \Delta_{SO}/3 \approx 114 \; \textrm{meV}$~\cite{Park2000:JAP,FariaJunior2012:JAP}. % FJ11
For the GaN barrier, 
$\Delta_1 = 10 \; \textrm{meV}$ and $\Delta_2 = \Delta_3 = 5.7 \; \textrm{meV}$,
and for the QW material $\textrm{In}_{0.1}\textrm{Ga}_{0.9}\textrm{N}$, 
$\Delta_1 = 13 \; \textrm{meV}$ and $\Delta_2 = \Delta_3 = 5.3 \; \textrm{meV}$. 
In the bulk case, such values of $\Delta_{1,2,3}$ provide an energy difference 
at the $\Gamma$-point in GaN of $\sim$5.1 meV for HH-LH and 21.9 meV for HH-CH. For 
$\textrm{In}_{0.1}\textrm{Ga}_{0.9}\textrm{N}$ the energy differences are $\sim$6 
meV and $\sim$22.9 meV for HH-LH and HH-CH, respectively. 
QW confinement and polarization fields can provide larger energy separations for 
the different wave functions (no nodes, one node, etc). However, the
typical HH-LH QW states separation with same number of nodes remains 
similar to the bulk energy values.
%GaN: HH-LH ~ 5.2 meV, HH-CH ~ 21.9 meV
%InGaN: HH-LH ~ 6 meV, HH-CH ~ 22.9 meV

The  resulting band structure of  $\textrm{In}_{0.1}\textrm{Ga}_{0.9}\textrm{N}$/GaN QW,
is presented in  Fig.~\ref{fig:bs_dos}(a),  showing the two confined 
conduction subbands, CB1 and CB2, and the top four confined valence subbands, 
HH1, LH1, HH2 and LH2, labeled according to the dominant component of the total 
envelope function~\cite{Chuang1996:APL}. 
Each subband is twofold degenerate in $\bm{k}=0$ and for nonzero $ {\bm k}$ values 
the effect of the asymmetric polarization field creates small spin splittings in 
the valence subbands~\cite{note:spinsplittings}, lifting Kramers degeneracy~\cite{Park2000:JAP}. 
Considering optical transitions at room temperature ($k_BT\sim$25 meV), the spin splittings are 
negligible as if the bands were twofold degenerate. Furthermore, because of the energy separation
of $\sim$150 (80) meV from CB1 (LH1) to CB2 (HH2) subbands, we can expect the emission range of the gain 
spectra to be ruled by CB1-HH1 ($2.963\; \textrm{eV}$) and CB1-LH1 ($2.973\; \textrm{eV}$) 
transitions. The corresponding density of states (DOS) shown in Fig.~\ref{fig:bs_dos}(b), 
confirms that spin-resolved DOS has equal contributions for spin up and spin down.

%===============================================================================

\section{III. Microscopic spin-dependent gain}
\label{sec:gain}

\begin{figure}[t]
\centering
\includegraphics*[width=8.5cm]{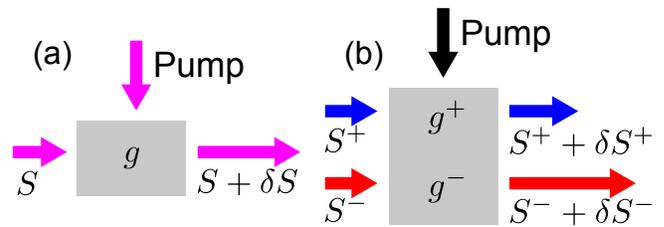}
\caption{Schematic of the optical gain, $g$,  for (a) conventional and (b) spin laser. 
With pumping/injection, a photon density $S$ increases by $\delta S$ 
as it passes across the gain region. In the spin laser this increase depends on 
the positive ($+$)/negative($-$) helicity of the light, $S=S^+ +S^-$~\cite{FariaJunior2015:PRB}.}
\label{fig:gain}
\end{figure}

Obtained electronic structure with the corresponding carrier populations provides the starting
point to microscopically calculate the optical gain depicted in Fig.~\ref{fig:gain}, the hallmark of lasers.  
The resulting gain coefficient (or gain spectrum) is the negative value of the absorption coefficient
and is calculated as~\cite{Haug:2004,Chuang:2009},
\begin{equation}
g^a_i(\omega)=C_0\underset{c,v, {\bm k}}{\sum} \left| p^a_{cv \bm{k}} \right|^2 \left(f_{c \bm{k}}-f_{v {\bm k}}\right)
\delta\left[\hbar\omega_{{cv} {\bm k}}-\hbar\omega\right],
\label{eq:epsI}
\end{equation}
where the summation indices $c$ 
and $v$ label the conduction and valence subbands, respectively
$p^a_{cv}$ is the interband dipole transition amplitude 
for the polarization of light $\alpha$, 
$f_{c(v) {\bm k}}$ is the Fermi-Dirac distribution for the electron occupancy in the conduction (valence)
subbands, $\hbar$ is the Planck's constant, $\omega_{{cv}\bm{k}}$ is the interband
transition frequency, and $\delta$ is the Dirac delta-function, which is often
replaced to include broadening effects for finite lifetimes~\cite{Chuang:2009,Chow:1999}. 
In the constant  $C_0 = 4\pi^2 e^2/(\varepsilon_0 c_l n_r m_0^2\omega\Omega)$, $\varepsilon_0$ is
the vacuum permittivity, $c_l$ is the speed of light (to distinguish it from the CB index),
$n_r$ is the dominant real part of the refractive index of the material, $e$ is the electron charge, 
$m_0$ is the free electron mass,   and $\Omega$ is the QW volume. 

Similar to ZB GaAs-based spin lasers~\cite{Holub2011:PRB,FariaJunior2015:PRB},
the dipole selection rules for the interband optical transitions are spin-conserving, 
i. e., % FJ10
the dipole matrix element does not change spin. 
%during the optical transition. % FJ10
Therefore, the gain coefficient for the light polarization $\alpha$ 
includes independent contributions of spin-up and spin-down carriers, 
\begin{equation}
g^a(\omega) = g^a_+(\omega) + g^a_-(\omega),
\label{eq:gain}
\end{equation}
denoted by the  subscripts $+$ and $-$, respectively.

\begin{figure}[t!]
\begin{center}
\includegraphics{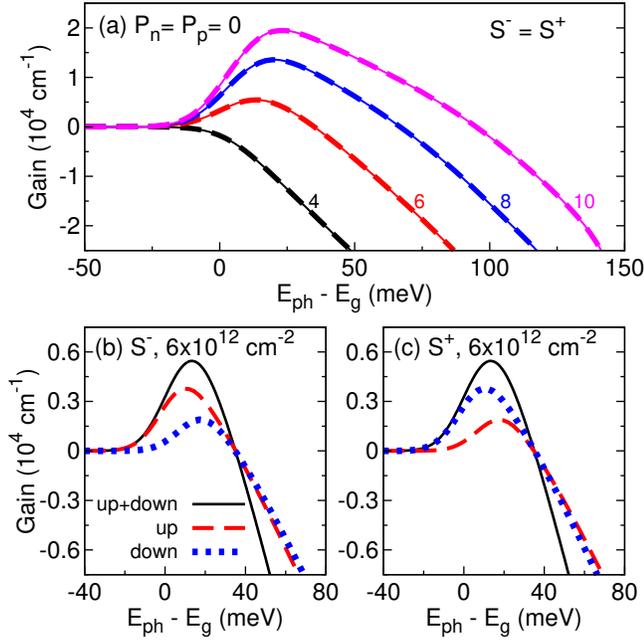}
\caption{(a) Gain spectra as a function of the photon energy  for spin-unpolarized electrons, $P_n=P_p=0$, 
at carrier densities $n=p=4,6,8,$ and $10 \times 10^{12} \; \textrm{cm}^{-2}$, labeling
the curves.  For these  values of  carrier densities, the emission is dominated by 
CB1-HH1 ($E_g$) and CB1-LH1 ($E_g$ + 10 meV) transitions. 
Spin-resolved gain coefficients for (b) $S^-$ and (c) $S^+$ at $n = p = 6 \times 10^{12} \; \textrm{cm}^{-2}$.}
\label{fig:gain_spin}
\end{center}
\end{figure}

To develop intuition and understand  the role of SOC in the optical transitions, we first illustrate the gain spectra 
on the example of conventional lasers.  This implies injecting vanishing electron and hole spin polarization, $P_n=P_p=0$, 
where the expression for $P_p$ is analogous to Eq.~(\ref{eq:Pn}). 
In Fig.~\ref{fig:gain_spin}(a)  we show such a gain spectra as function of photon energy for 
various  carrier densities. 

For calculated gain spectra in (In,Ga)N QWs it is customary to include various broadening effects.
In addition to the homogeneous broadening, frequently used in ZB QWs~\cite{Chow:1999,FariaJunior2015:PRB},
parametrized here by $sech$ with 10 meV full width at half-maximum (FWHM), we also consider an
inhomogeneous Gaussian broadening,  attributed to compositional and potential fluctuations.  Our choice of  Gaussian 
broadening with 20 meV FWHM  is consistent with a decreased broadening for smaller emission wavelengths 
in (In,Ga)N QW lasers and reported values relevant for wavelengths of $\sim$415 nm~\cite{Funato2013:APE} 
which corresponds to the typical energy of the gain peak in our calculations.

Because of the  broadening effects, the individual CB1-HH1 and CB1-LH1 
transitions that dominate the gain spectra cannot be distinguished [HH1 and LH1 
are 10 meV apart, see Fig.~\ref{fig:bs_dos}(a)]. On the other hand, 
by analyzing the spin-resolved gain we can identify  different contributions 
of CB1-HH1 and CB1-LH1 transitions. In Figs.~\ref{fig:gain_spin}(b) and \ref{fig:gain_spin}(c) 
we show the gain spectra decomposed in spin up and spin down transitions at 
$n = 6 \times 10^{12} \; \textrm{cm}^{-2}$ for $S^+$ and $S^-$ light polarization, respectively.
For the total gain we have $g^+=g^-$ which requires $g^+_-=g^-_+$ and $g^+_+=g^-_-$~\cite{FariaJunior2015:PRB},
as could be seen in Figs.~\ref{fig:gain_spin}(b) and \ref{fig:gain_spin}(c). 
Due to the small SOC energy in nitrides, the $S^-$ ($S^+$) gain peak of spin %I9
up (down) CB1-HH1 transition is twice as large as the spin down (up) CB1-LH1 transition. 
For a larger SOC energy, this ratio would increase. For example, in ZB 
GaAs spin laser~\cite{FariaJunior2015:PRB}, this ratio is $\sim$6 
(for a SOC energy of $\Delta_2 = \Delta_3 \approx$ 114 meV, compared to $\Delta_2 = \Delta_3 \sim 10-20$ meV in nitrides). % FJ11

\begin{figure}[t!]
\begin{center}
\includegraphics{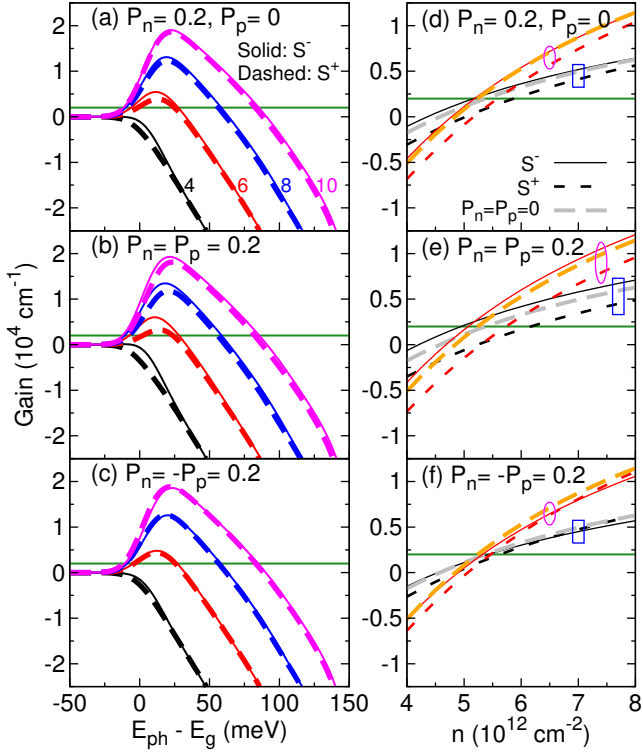}
\caption{Gain spectra 
for different spin polarizations (a) $P_n = 0.2$, $P_p = 0$, (b) $P_n = P_p = 0.2$, and (c) $P_n = -P_p = 0.2$.
Carrier densities $n=p=4,...,10 \times 10^{12} \; \textrm{cm}^{-2}$, label
the curves in (a). %I12
Gain values as function of carrier density for 
(d) $P_n = 0.2, P_p = 0$,
(e) $P_n = P_p = 0.2$, and (f) $P_n = -P_p = 0.2$ assuming photon energies at CB1-HH1
(curves indicated by the rectangle) and CB1-LH1 (curves indicated by the ellipse).
%transition energies. 
The solid horizontal line: gain threshold $g_{th} = 2 \times 10^{3} \; \textrm{cm}^{-1}$.
%As in Fig.~\ref{fig:gain_spin}(a), 
%The labeling in Fig.~\ref{fig:gain_spectra}(a) follows the scheme of Fig.~\ref{fig:gain_spin}(a). %FJ11
}
\label{fig:gain_spectra}
\end{center}
\end{figure}
% things to mention about the figure:
% 1) nonzero spin polarization the gain of S^+ and S^- are different
% 2) For Pn = -Pp the gain is also different
% 3) fixing photon energy ... square CB1-HH1 and ellipse CB1-LH1

We next turn to the gain properties in spin-lasers where injected carriers are spin-polarized.
Guided by the typical spin dynamics for ZB semiconductors in which hole spin relaxes 
nearly instantaneously, 
previous studies have % FJ10
largely focussed on spin lasers
with  nonzero $P_n$, but vanishing $P_p$. However, since the degeneracy of HH and LH 
in bulk WZ semiconductors is lifted by the crystal field potential, the spin relaxation times of holes 
in GaN could be comparable to those  of electrons~\cite{Brimont2009:JAP}. This is in stark contrast 
to bulk GaAs where at 300 K the hole spin relaxation time is three to four orders of 
magnitude shorter than for electrons~\cite{Zutic2004:RMP}.  We therefore also 
consider the effect of nonzero $P_p$, excluded in the two prior microscopic studies
of  gain spectra in spin lasers~\cite{Holub2011:PRB,FariaJunior2015:PRB}.

The gain for  WZ spin  lasers is shown in Fig.~\ref{fig:gain_spectra} as a function of  photon energy 
and carrier density.  These results confirm that the gain becomes helicity-dependent, $g^+\neq g^-$,
as %I9
known from ZB spin lasers. However, the role of simultaneous presence of 
nonvanishing $P_n$ and $P_p$ requires further attention. With fixed $P_n=0.2$ we see that
a change from $P_p=0$ to $P_p=0.2$ [panels (a) and (b)] enhances the difference between
the gain contribution for $S^-$ and $S^+$ , while a change from $P_p=0$ to $P_p=-0.2$  
[panels (a) and (c)] reduces such a difference. 

Since equal but opposite
electron spin polarizations [$P_n=-P_p$, Fig.~\ref{fig:gain_spectra}(c)] describe the vanishing total spin 
in the gain region, it is helpful to note another realization of a vanishing total spin 
in Fig.~\ref{fig:gain_spin}(a). Nevertheless, the gain spectra in these two cases are slightly different
which can be attributed to the different features of %FJ10
CB and VB  including their curvature, 
number of confined bands, and DOS. 
Thus, % FJ10
the difference between the gain contribution for $S^-$ and $S^+$ cannot be eliminated for $P_n=-P_p$. 

A complementary information about the calculated gain is given with its density dependence
in Figs.~\ref{fig:gain_spectra}(d)-(f). The results are shown for photon energies,
corresponding to the CB1-HH1 and CB1-LH1 transitions [recall Fig.~\ref{fig:bs_dos}] which can be 
individually favored by the cavity design in a single-mode VCSEL~\cite{Michalzik:2013}.
Several trends can be inferred.  For example,  a nonlinear gain-dependence on density is different for the 
two photon energies. With an increased carrier density,  CB1-LH1 transition provides larger gain values  %I9
than as compared to CB1-HH1. While a reference curve (long dashed)  for the gain of a conventional laser 
is lower than $g^-$ %I9
for $P_p=0$ and $P_p=0.2$ [panels (d) and (e)], the situation is 
reversed above the gain threshold (green horizontal line) for $P_p=-0.2$ [panel (f)] where at larger
density $g^+ > g^-$ is possible.%I9

\begin{figure}[h!]
\begin{center}
\includegraphics{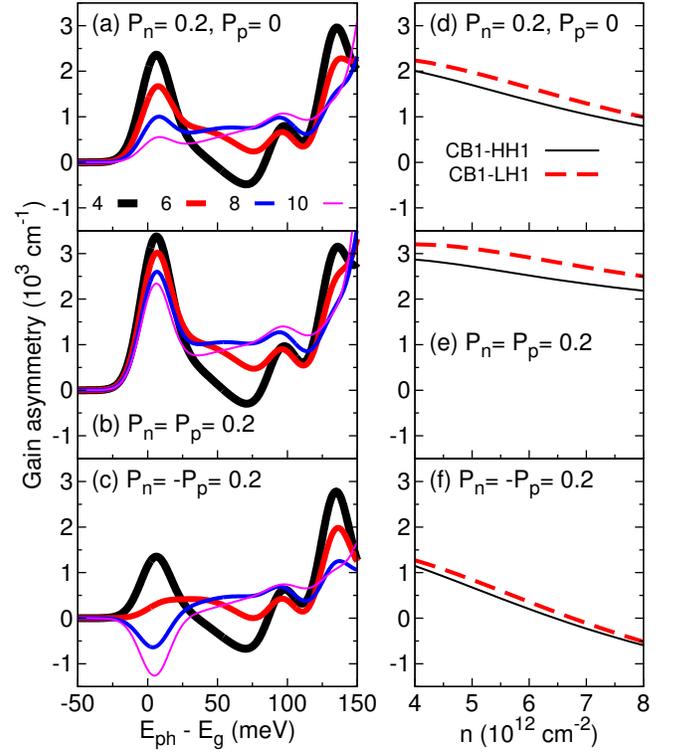}
\caption{Gain asymmetry (a)-(c) as a function of photon energy and (d)-(f) carrier density for the data from Fig.~\ref{fig:gain_spectra}.}
\label{fig:gain_asy}
\end{center}
\end{figure}

To better understand the helicity-dependent gain, it is useful to calculate the corresponding
gain asymmetry,
\begin{equation}
g_\text{asy}(\omega) = g^-(\omega)- g^+(\omega).
\label{eq:asy}
\end{equation}
an important figure of merit in spin lasers.
Considering that lasers have nonlinear light-injection characteristics, such gain asymmetry
could enable robust spin filtering or spin amplification~\cite{Gothgen2008:APL}. Close to the lasing threshold even 
a small carrier spin polarization in the gain region can lead to completely circularly polarized
emitted light~\cite{Iba2011:APL}.  

The results for the gain asymmetry,  extracted from Fig.~\ref{fig:gain_spectra}, 
are shown in Fig.~\ref{fig:gain_asy} as a function of the photon energy and the carrier density.
While a large $|g_\text{asy}|$ is desirable, it is crucial that it corresponds to the $g>0$ regime.
For example, the largest $|g_\text{asy}|$ in Figs.~\ref{fig:gain_spectra}(a) and (c) is found
for photon energies of 125 - 140 meV above the band gap. However, as seen in 
Figs.~\ref{fig:gain_spectra}(a) and (c), this range corresponds to the absorption regime ($g<0$)
and such $g_\text{asy}$  does not influence the emitted light.  As $P_p$ and $n$ vary,
the largest useful $|g_\text{asy}|$ is found slightly above the gap.  
As shown in Figs.~\ref{fig:gain_asy}(d)-(f), to enhance $|g_\text{asy}|$ a lower density 
and CB1-LH1 are slightly better. An interesting deviation from these trends is seen
in Fig.~\ref{fig:gain_asy} for $P_n = -P_p = 0.2$.  Near the band gap, an increase in $n$ 
leads to the sign change of $g_\text{asy}$ and its  maximum magnitude in the $g<0$ 
regime for the larges shown carrier density. This behavior points to yet unexplored opportunities
to optimize the operation of spin lasers with a simultaneous spin polarization of electrons and holes.

Our results show that despite the small SOC energy of WZ nitrides, considered
detrimental for optical spin injection, the gain asymmetry remains robust. Another  
important figure of merit of spin lasers is their threshold reduction, the lasing 
operation %FJ10
could 
be attained at lower injected carrier density  than in conventional lasers.
We will analyze this behavior in the next section.  
%===============================================================================

\section{IV. Rate equations}
\label{sec:RE}

Here we briefly review a complementary approach based on rate equations (REs) and
discuss how its understanding can be enhanced from our microscopic gain calculations.
REs have been successfully used to describe both conventional and 
spin lasers~\cite{Chuang:2009,Michalzik:2013,Coldren:2012,Zutic:2011}.
An advantage of this approach is its simplicity. REs can provide a direct relation
between material characteristics and device parameters, as well as often allowing
analytical solutions and an effective method to elucidate many trends in the operation
of lasers~\cite{Chuang:2009,Coldren:2012}. With notation widely used for conventional 
lasers~\cite{Chuang:2009,Coldren:2012}, generalized to include spin- and helicity-resolved
quantities, we can write REs as~\cite{Lee2014:APL,Wasner2015:APL},
\begin{widetext}
\begin{eqnarray}
\frac{dn_{\pm}}{dt} = J^n_{\pm}-g_{\pm}(n_{\pm},p_{\pm},S)S^{\mp}-(n_{\pm}-n_{\mp})/\tau_{sn}-R_{sp}^{\pm} \\ 
\label{eq:ren}
\frac{dp_{\pm}}{dt} = J^p_{\pm}-g_{\pm}(n_{\pm},p_{\pm},S)S^{\mp}-(p_{\pm}-p_{\mp})/\tau_{sp}-R_{sp}^{\pm} \\ 
\label{eq:rep}
\frac{dS^{\pm}}{dt} = \Gamma g_{\mp}(n_{\mp},p_{\mp},S)S^{\pm}-S^{\pm}/\tau_{ph}+\beta \Gamma R_{sp}^{\mp}. 
\label{eq:reS}
\end{eqnarray}
\end{widetext}

In the gain term, $g_\pm(n_\pm,p_\pm,S)= g_0(n_\pm+p_\pm-n_{\mathrm{tran}})/(1+\epsilon S)$,
$n_{\mathrm{tran}}$ is the transparency density, and $\epsilon$ is the gain compression
factor~\cite{Chuang:2009,Coldren:2012}, ensuring that the output light $S$ does not increase indefinitely
with injection $J$, $g_0$ is the gain parameter, and $\Gamma$ is the optical confinement factor. The electron
spin relaxation is given by $(n_\pm-n_\mp) /\tau_{sn}$, where $\tau_{sn}$ is the electron spin relaxation time 
($\tau_{sp}$ for holes)~\cite{Zutic2003:APL}.  %I13, I10
The carrier recombination $R_{\mathrm{sp}}^\pm$ can have various dependences on carrier density~\cite{Gothgen2008:APL}
and be characterized by a carrier recombination time $\tau_r$. $\beta$ is the
fraction of the spontaneous recombination producing light  coupled to the resonant
cavity, and $\tau_{\mathrm ph}$ is the photon lifetime, to model optical losses~\cite{Rudolph2003:APL,Holub2007:PRL,Gothgen2008:APL}.

\begin{figure}[h]
\begin{center}
\includegraphics{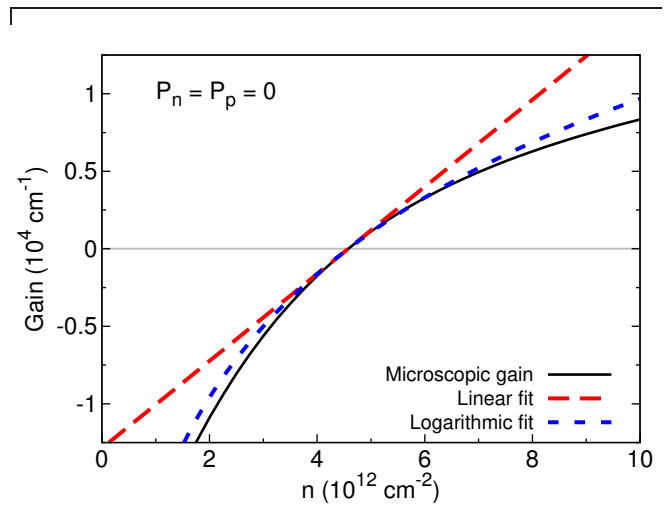}
\caption{Calculated gain from the $\bm{k{\cdot}p}$ method 
[solid curve, results from Fig.~\ref{fig:gain_spectra}(d)] %I12 
fit to
the linear (long dashed) and logarithmic (short dashed) density dependence by matching
the zero gain value.} 
\label{fig:fit}
\end{center}
\end{figure}

While the $\bm{k{\cdot}p}$ method does not include spin relaxation ($\tau_{sn}/\tau_r$, $\tau_{sp}/\tau_r \rightarrow \infty$),
similar dynamical effects are easily included in REs. However, REs rely on various input parameters that can be obtained
from experiments or microscopic calculations. A more complete description of spin lasers can be therefore developed by
combining the  $\bm{k{\cdot}p}$ method and the macroscopic RE model. We illustrate this approach
by focusing on the optical gain in WZ spin lasers. Specifically, the gain parameter %I10
 and the transparency density in 
the gain term in REs, can be obtained by fitting, for each $P_n$ and $P_p$, %I9 
the carrier density dependence of the calculated microscopic gain
%I12 [the data 
presented in Figs.~\ref{fig:gain_spectra}(d)-(f). %I12]. % FJ11 
% calculated by the $\bm{k{\cdot}p}$  method. 

Following the REs for spin lasers~\cite{Rudolph2003:APL,Holub2007:PRL,Gothgen2008:APL,Lee2014:APL,Wasner2015:APL} 
we use a simple linear dependence of gain on the carrier density to 
provide a better comparison with the published work.
This is illustrated in Fig.~\ref{fig:fit} for calculated gain of a conventional WZ  laser. 
In the linear fit, the slope of the gain at $n_{\mathrm{tran}}$ (where $g=0$) in REs is matched with the slope of the calculated gain.
However, we note that the logarithmic gain model, often used in conventional QW lasers~\cite{Chuang:2009},
would be a better fit. Another difference between REs and the calculated gain is the  
helicity-dependent gain coefficient (recall Figs.~\ref{fig:gain_spectra} and ~\ref{fig:gain_asy}) 
and we include that behavior by fitting the RE gain for each helicity separately. 
To follow the $\bm{k{\cdot}p}$  method we choose $\tau_{sn}/\tau_r$, $\tau_{sp}/\tau_r \gg 1$, 
rather than seeking the best possible fit between the two methods.  Likewise, we choose
$\epsilon=0$ even though the gain compression could give a better agreement  at larger $n$.
The remaining RE parameters are assigned 
from the previous work~\cite{Arif2008:IEEEJQE}. %I9 not complete, cavity parameters not in that work

\begin{figure}[t]
\begin{center}
\includegraphics{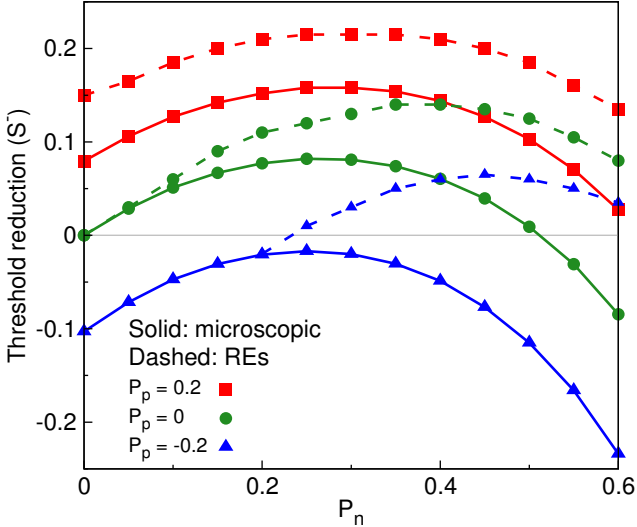}
\caption{Comparing the threshold reduction for $S^-$ light polarization for different 
hole spin polarizations, $P_p=0, \pm 0.2$, between the  $\bm{k{\cdot}p}$ method (solid) 
and REs (dashed curves).} 
\label{fig:TR}
\end{center}
\end{figure}

Unlike a single lasing threshold, $J_T$, in conventional lasers, with 
spin-dependent gain, there are two lasing thresholds in spin lasers, 
$J_{T1} \leq J_{T2}$ which delimits three operating regimes~\cite{Gothgen2008:APL} :
(i) For  $J \leq J_{T1}$ a spin LED regime,
(ii) For  $J_{T1} \leq J \leq J_{T2}$ there is a spin-filtering regime and the lasing
with only one helicity. (iii) For $J\geq J_{T2}$, there is a lasing with both helicities.
It is then convenient to define the threshold reduction,
\begin{equation}
r=1-J_{T_1}/J_T,
\label{eq:r}
\end{equation}
as an important  figure of merit that influences both the steady-state and dynamical operation of 
spin lasers~\cite{Lee2010:APL}. 
 
In Fig.~\ref{fig:TR}  we compare the threshold reduction as a function of electron spin polarization
calculated using the microscopic method and REs, for several values of hole spin polarization.
Despite noticeable differences between the two methods,  they both show an overall  
{\em nonmonotonic}  dependence of  $r$ on $P_n$, preserved for each $P_p$. It is instructive to note
that previously studied REs with $P_p=0$ and fixed $g_0$ yield a {\em monotonic} increase 
in $r$ with $P_n$, from $r=0$ at $P_n=0$ to $r=1/2$ at $P_n=1$~\cite{Gothgen2008:APL}.
However, %I9
using REs with a linear fit of the gain for  $P_p=0$ at  each $P_n$ shows in Fig.~\ref{fig:TR} 
a much closer agreement  to the microscopic gain %I9
results and, by constructions, the two methods coincide 
at $P_n=P_p=0$.
 
Including the hole spin polarization, the disagreement between the two methods is more 
pronounced for $P_p=-0.2$, than for $P_p=0.2$.  The corresponding RE results largely
fail to capture the calculated threshold increase ($r<0$, reported also in ZB lasers~\cite{Holub2011:PRB,FariaJunior2015:PRB}) 
and are not properly defined for $P_n<0.2$.
To explore why the RE results for $P_p=-0.2$ are worse, it is useful to recall the dipole optical selection rules
for transitions sketched in Fig.~\ref{fig:ZBWZpol}. In our notation that means that both spontaneous
and stimulated recombination (optical gain) involve only electrons and holes of the same spin. For example,
spontaneous radiative recombination has terms $n_+p_+$ or $n_-p_-$~\cite{note:recombine}. %I9
However, in Eqs.~(\ref{eq:ren}) or (\ref{eq:reS}) 
the gain term does not accurately respect these  selection rules. For a sufficiently large carrier 
density  the lasing would occur, even if the carrier spins are not compatible with the selection rules. 
When $P_n$ and $P_p$ have the opposite sign there are more carriers having a ``wrong spin'' to satisfy
the selection rules leading to a worse agreement with the  microscopic results. Such a disagreement 
would be less pronounced for shorter spin relaxation times, allowing ``wrong spin carriers'' to recombine 
while respecting the selection rules. 

It is also possible to address  the missing RE data for $P_n<0.2$.
In the steady-state,  Eq.~(\ref{eq:reS}) implies, 
\begin{equation}
n_{\mp}+p_{\mp}= n_{tran} +1/ (\Gamma g_0 \tau_{ph}) - \beta R^{\mp}_{sp}/(g_0 S^{\pm}).
\label{eq:steady}
\end{equation}
In the operating regime (iii): $J>J_{T2}$, both 
helicities lase and $S^{\pm}$ are large,
which yields,
$n_{\mp}+p_{\mp}\approx n_{tran} + 1/(\Gamma g_0 \tau_{ph})$,
Therefore, $n_{+}+p_{+}\approx n_{-}+p_{-}$. Together with the charge neutrality, we have
$p_+=n_-$  and $p_-=n_+$, which means that $P_n=-P_p$ is guaranteed in the regime (iii). This is relevant
for the case $P_p=-0.2$ and $P_n<0.2$, because emitted $S^-$ is associated with minority 
instead of majority spin, such that the lasing of $S^-$ is in the regime (iii). 
The required $P_n=-P_p$ in the third regime thus reduces the freedom of a realizable 
spin polarization in REs.

%===============================================================================

\section{V. Conclusions}
\label{sec:conclusion} 

Our framework of combining microscopic gain calculations with simple rate equations provides predictive and
computationally inexpensive materials-specific approach to explore spin lasers.  The choice of wurtzite lasers was 
guided by the first realization of an electrically manipulated %I9 pumped 
spin laser at room temperature~\cite{Chen2014:NN} and the 
absence of any prior theoretical work. In contrast to zinc-blende GaAs, in wurtzite GaN there is a much smaller spin-orbit 
coupling, usually considered as a detrimental situation for optical spin injection. We have shown that even such 
a small spin-orbit coupling in wurtzites 
yields robust signatures of a spin-dependent gain, including the gain asymmetry, desirable for spin lasers.  
%I13
With the %simultaneous 
presence of  nonvanishing electron and hole spin polarization,  largely overlooked in the previous studies,
the gain asymmetry can even change its sign by simply increasing the carrier density. The lasing threshold reduction has 
a nonmonotonic depenedence on electron spin polarization, even for a nonvanishing hole spin polarization. 

While  a weak spin-orbit coupling is expected to lead to an enhanced spin relaxation times, this is not the case for GaN 
which has a defect dominated spin-relaxation and electron spin relaxation times about an order of magnitude shorter 
than in GaAs~\cite{Brimont2009:JAP,Bhattacharya2016:APL}.  Although materials advances could enable longer spin relaxation times 
in GaN,  the current values are already suitable for digital and high-frequency operation of spin  
lasers~\cite{Lee2014:APL,Wasner2015:APL}. 

The present framework can be adapted for other materials and laser geometries. With an increasing interest 
in non-nitride III-V wurtzite materials with large spin-orbit 
coupling~\cite{De2010:PRB,Cheiwchanchamnangij2011:PRB,FariaJunior2016:PRB,Gmitra2016:PRB}, 
we expect they could facilitate optically-injected spin lasers at room temperature. While we have focused
on spin VCSELs, our approach would also be useful  for vertical external  cavity surface emitting lasers 
(VECSELs)\cite{Frougier2015:OE,Frougier2013:APL}.  They enable depositing a thin-film ferromagnet 
just 100-200 nm away from  the gain region for spin injection at the room temperature. 
Various spin and phonon lasers 
can also be implemented using intraband transitions within the conduction band~\cite{Khaetskii2013:PRL} or in metallic 
systems~\cite{Korenivski2013:EPL,Naidyuk2012:NJP}. It would be interesting to develop a suitable description 
for them by combining microscopic gain calculations and simple rate equations.

%I13
An important materials challenge for the %future 
advances in wurtzite spin lasers would be to establish magnetic 
regions and their detailed theoretical description for robust electrical spin injection into the gain region. 
In zinc-blende semiconductors a number of such materials are already 
available~\cite{Zutic2004:RMP,Fabian2007:APS,Hanbicki2002:APL,Zega2006:PRL,Salis2005:APL}. In addition 
to demonstrating that Fe$_3$O$_4$ nanomagnets are suitable for wurtzite spin lasers~\cite{Chen2014:NN}, 
many other opportunities could be explored. For example, ferromagnetic semiconductors provide 
electrically- and optically-controlled magnetic
properties~\cite{Zutic2004:RMP,Koshihara1997:PRL,Ohno2000:N,Dietl2014:RMP, DiasCabral2011:PRB}, while supporting ultrafast optical 
processes~\cite{Wang2006:JPCM}. 
With a thin barrier region, even simple ferromagnets 
may enable tunable carrier spin polarization relying on gate-controlled magnetic proximity effects~\cite{Lazic2016:PRB}.

We thank N.~C. Gerhardt for valuable discussions of the optical gain.
This work was supported by  
NSF ECCS-1508873, U.S. ONR N000141310754,
FAPESP Grants No. 2012/05618-0, 2011/19333-4,
CNPq Grants No. 304289/2015-9, 246549/2012-2,  
CAPES-CsF 88887.125287/2015-00, and 
MOST-105-2811-M-147, Taiwan, Republic of China.

%===============================================================================

%===============================================================================


\begin{thebibliography}{99}
\bibitem{Hallstein1997:PRB}
S. Hallstein, J.~D. Berger, M. Hilpert, H.~C. Schneider, W.~W. R\"{u}hle, F. Jahnke, 
S.~W. Koch, H.~M. Gibbs, G. Khitrova, and M. Oestreich,
Phys. Rev. B {\bf 56}, R7076 (1997).

\bibitem{Ando1998:APL}
H. Ando, T. Sogawa, and H. Gotoh, 
Appl. Phys. Lett. {\bf 73}, 566 (1998). 

\bibitem{Rudolph2003:APL}
J. Rudolph, D. H\"{a}gele, H.~M. Gibbs, G. Khitrova, and M. Oestreich, 
Appl. Phys. Lett. {\bf 82}, 4516 (2003);
J. Rudolph, S. D\"{o}hrmann, D.~H\"{a}gele, M. Oestreich, and W. Stolz,
Appl. Phys. Lett. {\bf 87}, 241117 (2005).

\bibitem{Holub2007:PRL}
M. Holub, J. Shin, D. Saha, and P. Bhattacharya, 
Phys. Rev. Lett. {\bf 98}, 146603 (2007).

\bibitem{Hovel2008:APL} 
S. H\"{o}vel, A. Bischoff, and N.~C. Gerhardt,   M.~R. Hofmann, and T. Ackemann,  A. Kroner, and R. Michalzik, Appl. Phys. Lett. {\bf 92}, 041118 (2008).

\bibitem{Basu2008:APL} 
D. Basu, D. Saha, C. C.Wu, M. Holub, Z. Mi, and P. Bhattacharya, 
Appl. Phys. Lett. {\bf 92}, 091119 (2008).

\bibitem{Basu2009:PRL}
D. Basu, D. Saha, and P. Bhattacharya,
Phys. Rev. Lett. {\bf 102}, 093904 (2009).

\bibitem{Saha2010:PRB}
D. Saha, D. Basu, and P. Bhattacharya, 
Phys. Rev. B {\bf 82}, 205309 (2010).

\bibitem{Fujino2009:APL}
H. Fujino, S. Koh, S. Iba, T. Fujimoto, and H. Kawaguchi, 
Appl. Phys. Lett. {\bf 94}, 131108 (2009).

\bibitem{Jahme2010:APL} 
M. Li, H. J{\"a}hme, H. Soldat, N.~C. Gerhardt, M.~R. Hofmann, and T. Ackemann,  A. Kroner, and R. Michalzik Appl. Phys. Lett. {\bf 97}, 191114 (2010).

\bibitem{Gerhardt2011:APL}
N. C. Gerhardt, M. Y. Li, H. J\"ahme, H. H\"opfner, T. Ackemann, and M. R. Hofmann,
Appl. Phys. Lett. {\bf 99}, 151107 (2011).

\bibitem{Iba2011:APL}
S. Iba, S. Koh, K. Ikeda, and H. Kawaguchi, 
Appl. Phys. Lett. {\bf 98}, 081113 (2011).

\bibitem{Frougier2013:APL}
J. Frougier, G. Baili, M. Alouini, I. Sagnes, H. Jaffr\`{e}s, A. Garnache, C. Deranlot, D. Dolfi, and J.-M. George,
Appl. Phys. Lett. {\bf 103}, 252402 (2013).

\bibitem{Frougier2015:OE}
J. Frougier, G. Baili, I. Sagnes, D. Dolfi, J.-M. George, and M. Alouini,
Opt. Expr. {\bf 23}, 9573 (2015).

\bibitem{Alharthi2015:APL}
S.~S. Alharthi, A. Hurtado, V.-M. Korpijarvi, M. Guina, I.~D. Henning, and M.~J. Adams,
Appl. Phys. Lett. {\bf 106}, 021117 (2015).

\bibitem{Alharthi2015:APLb}
S.~S. Alharthi, J. Orchard, E. Clarke,  I.~D. Henning, and M.~J. Adams,
Appl. Phys. Lett. {\bf 107}, 151109 (2015).

\bibitem{Hsu2015:PRB}
F.-k. Hsu, W. Xie,  Y.-S. Lee, S.-D. Lin, and C.-W. Lai,
Phys. Rev. B {\bf 91}, 195312 (2015).

\bibitem{Sinova2012:NM}
J. Sinova and I. \v{Z}uti\'c, 
Nature Mater. {\bf 11}, 368 (2012).

\bibitem{Lee2010:APL}
J. Lee, W. Falls, R. Oszwa\l dowski, and I. \v{Z}uti\'c,
Appl. Phys. Lett. {\bf 97}, 041116 (2010).

\bibitem{Boeris2012:APL}
G. Bo\'{e}ris, J. Lee, K. V\'{y}born\'{y}, and I. \v{Z}uti\'{c},
Appl. Phys. Lett. {\bf 100}, 121111 (2012).

\bibitem{Lee2012:PRB}
J. Lee, R. Oszwa{\l}dowski, C. G{\o}thgen, and I. \v{Z}uti\'{c},
Phys. Rev. B {\bf 85}, 045314 (2012).

\bibitem{Lee2014:APL}
J. Lee, S. Bearden, E. Wasner, and I. \v{Z}uti\'{c},
Appl. Phys. Lett. {\bf 105}, 042411 (2014).

\bibitem{Wasner2015:APL}
E. Wasner, S. Bearden, J. Lee, and I. \v{Z}uti\'{c},
Appl. Phys. Lett. {\bf 107},  082406 (2015).

\bibitem{FariaJunior2015:PRB}
P.~E. Faria~Junior, G. Xu, J. Lee, N.~C. Gerhardt, G.~M. Sipahi, and I. \v{Z}uti\'c,
Phys. Rev. B {\bf 92}, 075311 (2015).

\bibitem{Hopfner2014:APL}
H. H\"opfner, M. Lindemann, N. C. Gerhardt, and M. R. Hofmann,
Appl. Phys. Lett. {\bf 104}, 022409 (2014).

\bibitem{Lindemann2016:APL}
M. Lindemann, T. Pusch, R. Michalzik, N.~C Gerhardt, and M.~R. Hofmann,
Appl. Phys. Lett. {\bf 108}, 042404 (2016).

\bibitem{Pusch2015:EL}
T. Pusch, M. Lindemann, N.~C. Gerhardt, M.~R. Hofmann, and R. Michalzik, 
Electron. Lett. {\bf 51}, 1600 (2015).

\bibitem{Meier:1984}
{\it Optical Orientation,} edited by F. Meier and B.~P. Zakharchenya
(North-Holland, New York, 1984).

\bibitem{Chen2014:NN}
J.-Y. Chen, T.-M. Wong, C.-W. Chang, C.-Y. Dong, and Y.-F Chen,
Nat. Nanotech. {\bf 9}, 845 (2014).

\bibitem{Zutic2014:NN}
I. \v{Z}uti\'c and P.~E. Faria~Junior,
Nat. Nanotech. {\bf 9}, 750 (2014).

\bibitem{Gothgen2008:APL}
C. G\o thgen, R. Oszwa\l dowski, A. Petrou, and I. \v{Z}uti\'c,
Appl. Phys. Lett. {\bf 93}, 042513 (2008).

\bibitem{Holub2011:PRB}
M. Holub and B.~T. Jonker,
Phys. Rev. B {\bf 83}, 125309 (2011).

\bibitem{Akasaki2015:RMP}
I. Akasaki, Rev. Mod. Phys. 87, 1119 (2015).

\bibitem{Amano2015:RMP}
H. Amano, Rev. Mod. Phys. 87, 1133 (2015).

\bibitem{Nakamura2015:RMP}
S. Nakamura, Rev. Mod. Phys. 87, 1139 (2015).

\bibitem{Chen2005:APL}
W.~M. Chen, I.~A. Buyanova, K. Nishibayashi,  Kayanuma, K. Seo, A. Murayama, Y. Oka,
G. Thaler, R. Frazier, C.~R. Abernathy, F. Ren, S.~J. Pearton, C.-C. Pan, G.-T. Chen and J.-I Chyi,
Appl. Phys. Lett. {\bf 87}, 192107 (2005). %I12 Nils corrected volume

\bibitem{Winkler:2003} %I13
R. Winkler,
{\em Spin-orbit Coupling Effects in Two-Dimensional Electron and Hole Systems},
(Springer, New York, 2003).

\bibitem{Lee2014:PRB}
J. Lee, K. V\'{y}born\'y, J.~E. Han, and I. \v{Z}uti\'c, 
Phys. Rev. B {\bf 89}, 045315 (2014).

\bibitem{Chuang1996:PRB}
S.~L. Chuang and C.~S. Chang,
Phys. Rev. B {\bf 54}, 2491 (1996).

\bibitem{Pierce1975:PLA}
D. Pierce, F. Meier and P. Z\"urcher,
Phys. Lett. A {\bf 51}, 465 (1975).

\bibitem{Zutic2004:RMP}
I. \v{Z}uti\'c, J. Fabian, and S. Das Sarma,
Rev. Mod. Phys. {\bf 76}, 323 (2004).

\bibitem{note:GaN}
In stark contrast to very dilute ZB nitrides with strong SOC,  such as Ga$_{0.67}$In$_{0.33}$N$_{0.016}$As$_{0.984}$ 
QW-based spin lasers, K. Schires, R.~A. Seyab, A. Hurtado, V.-M.  Korpij\"{a}rvi, M. Guina, I.~D. Henning, and M.~J. Adams,
Opt. Expr. {\bf 20}, 3550 (2012).

\bibitem{Banerjee2013:APL}
D. Banerjee, R. Adari, S. Sankaranarayan, A. Kumar, S. Ganguly, R.~W. Aldhaheri, M.~A. Hussain, A.~S. Balamesh, and D. Saha,
Appl. Phys. Lett. {\bf 103}, 242408 (2013). %I2 Nils fixed vol., page, year

\bibitem{Chen2010:APL}
L.-C. Chen, C.-H. Tien, and C.-S. Mu, 
%Effects of spin-polarized injection and photoionization of MnZnO film on GaN-based light emitting diodes
Opt. Expr. {\bf 18}, 2302 (2010). %I12 Nils suggested 

\bibitem{Chen2014:NL}
J. Y. Chen, C. Y. Ho, M. L. Lu,  L. J. Chu, K. C. Chen, S. W. Chu, W. Chen, C. Y. Mou, and Y.-F. Chen,
Nano Lett. {\bf 14}, 3130 (2014).

\bibitem{note:pols}
See,  for example, Chapter 3 of {\it Nitride Semiconductor Devices: Principles and Simulation}, edited by J. Piprek 
(Wiley, Weinheim, 2007).

\bibitem{Chuang:2009}
S.~L. Chuang,
{\it Physics of Optoelectronic Devices},
2$^\textrm{nd}$ Edition (Wiley, New York, 2009).

\bibitem{FariaJunior2012:JAP}
P.~E. Faria~Junior and G.~M. Sipahi,
J. Appl. Phys. {\bf 112}, 103716 (2012).

\bibitem{Miao2012:PRL}
M.~S. Miao, Q. Yan, C.~G. Van de Walle, W.~K. Lou, L.~L. Li, and K. Chang, %I13 g added
Phys. Rev. Lett. {\bf 109}, 186803 (2012).

\bibitem{FariaJunior2014:JAP}
P.~E. Faria~Junior, T. Campos, and G.~M. Sipahi,
J. Appl. Phys. {\bf 116}, 193501 (2014).

\bibitem{Sipahi1996:PRB}
G.~M. Sipahi, R. Enderlein, L.~M.~R. Scolfaro and J.~R. Leite,
Phys. Rev. B {\bf 53}, 9930 (1996).

\bibitem{Lu2008:APL}
T.-C. Lu, C.-C. Kao, H.-C. Kuo, G.-S. Huang and S.-C. Wang,
Appl. Phys. Lett. {\bf 92}, 141102 (2008).

\bibitem{Lu2010:APL}
T.-C. Lu, S.-W. Chen, T.-T. Wu, P.-M. Tu, C.-K. Chen, C.-H. Chen, Z.-Y. Li, H.-C. Kuo and S.-C. Wang,
Appl. Phys. Lett. {\bf 97}, 071114 (2010).

\bibitem{Kasahara2011:APE}
D. Kasahara, D. Morita, T. Kosugi, K. Nakagawa, J. Kawamata, Y. Higuchi, H. Matsumura and T. Mukai,
Appl. Phys. Expr. {\bf 4}, 072103 (2011).

\bibitem{Lin2014:LPL}
B.~C. Lin, Y.~A. Chang, K.~J. Chen, C.~H. Chiu, Z.~Y. Li, Y.~P. Lan, C.~C. Lin,
P.~T. Lee, Y.~K. Kuo, M.~H. Shih, H.~C. Kuo, T.~C. Lu and S.~C. Wang,
Laser Phys. Lett. {\bf 11}, 085002 (2014).

\bibitem{Vurgaftman2001:JAP}
I. Vurgaftman, J.~R. Meyer and L.~R. Ram-Mohan, J. Appl. Phys. {\bf 89}, 5815 (2001).

\bibitem{Dong2013:JAP}
L. Dong, J. V. Mantese, V. Avrutin, \"U \"Ozg\"ur, H. Morko\c{c} 
and S. P. Alpay, J. Appl. Phys. {\bf 114}, 043715 (2013).

\bibitem{note:web}
http://www.ioffe.ru/SVA/NSM/Semicond/

\bibitem{Chuang1996:APL}
S.~L. Chuang and C.~S. Chang, Appl. Phys. Lett. {\bf 68}, 1657 (1996).

\bibitem{note:spinsplittings}
The VB spin splitting appears because of the off-diagonal SOC energy
$\Delta_3$ in the Hamiltonian that mediates the coupling between the asymmetric
potential and $k$-dependent terms. Within this $\bm{k{\cdot}p}$ model the CB has
no SOC interaction with other bands, therefore the spin splitting is not seen in
the CB structure.

\bibitem{Park2000:JAP}
S.~H. Park and S.~L. Chuang,
J. Appl. Phys. {\bf 87}, 353 (2000).

\bibitem{Haug:2004}
H. Haug and S.~W. Koch,
{\it Quantum Theory of Optical and Electronic Properties of Semiconductors},
4$^\textrm{th}$ Edition (World Scientific Publishing, Singapore, 2004).

\bibitem{Chow:1999}
W.~W. Chow and S.~W. Koch,
{\it Semiconductor-Laser Fundamentals: Physics of the Gain Materials},
(Springer, New York, 1999).

\bibitem{Funato2013:APE}
M. Funato, Y.~S. Kim, Y. Ochi, A. Kaneta, Y. Kawakami, T. Miyoshi  and S. Nagahama, 
Appl. Phys. Express {\bf 6}, 122704 (2013).

\bibitem{Brimont2009:JAP}
C.  Brimont, M. Gallart, A. Gadalla, Olivier Cr\'{e}gut, Bernd H\"{o}nerlage, and P. Gilliot,
J. Appl. Phys. {\bf 105}, 023502 (2009).

\bibitem{Michalzik:2013}
{\it VCSELs Fundamentals, Technology and Applications of Vertical-Cavity Surface-Emitting Lasers},
edited by R. Michalzik (Springer, Berlin, 2013).

\bibitem{Coldren:2012}
L.~A. Coldren, S.~W. Corzine, and M.~L. Ma\v{s}ovi\'{c},
{\it Diode Lasers and Photonic Integrated Circuits,} 2$^\textrm{nd}$ Edition
(Wiley, Hoboken, 2012).

\bibitem{Zutic:2011}
I. \v{Z}uti\'{c}, R. Oszwa\l dowski,  C. G\o thgen, and J. Lee, Semiconductor Spin-Lasers, in
Handbook of Spin Transport and Magnetism, edited by E. Y. Tsymbal and I. \v{Z}uti\'{c}
(CRC Press/Taylor\&Francis, New York, 2011), pp. 731-746.

\bibitem{Zutic2003:APL}
I. \v{Z}uti\'c, J. Fabian, and S. Das Sarma,
Appl. Phys. Lett. {\bf 82}, 221 (2003).

\bibitem{Arif2008:IEEEJQE}
R. A. Arif, H. Zhao, Y.-K. Ee, and N. Tansu,  IEEE J. Quant. Electron. {\bf 44},  573 (2008).

\bibitem{note:recombine}
Within the rate-equation description, including a widely-used spin-flip model
[M. San Miguel, Q. Feng, and J.V. Moloney, Phys. Rev. A {\bf 52}, 1728 (1995)],
the $n_+p_+$ recombination only gives $S^-$ helicity of the emitted light.
This means that there is only one type of holes within  a four-band model 
(CB and VB with a twofold spin degeneracy).

\bibitem{Bhattacharya2016:APL}
A. Bhattacharya, M.~Z. Baten, and P. Bhattacharya,
Appl. Phys. Lett. {\bf 108}, 042406 (2016).

\bibitem{De2010:PRB}
A. De and C.~E. Pryor,
Phys. Rev. B {\bf 81}, 155210 (2010).

\bibitem{Cheiwchanchamnangij2011:PRB}
T. Cheiwchanchamnangij and W.~R. Lambrecht,
Phys. Rev. B {\bf 84}, 035203 (2011).

\bibitem{FariaJunior2016:PRB}
P.~E. Faria~Junior, T. Campos, C.~M.~O. Bastos, M. Gmitra, J. Fabian and G.~M. Sipahi,
Phys. Rev. B {\bf 93}, 235204 (2016).

\bibitem{Gmitra2016:PRB}
M. Gmitra and J. Fabian, Phys. Rev. B {\bf 94}, 165202 (2016).
\bibitem{Khaetskii2013:PRL} 
A. Khaetskii, V.~N. Golovach, X. Hu,  and I. \v{Z}uti\'c,
Phys. Rev. Lett. {\bf 111}, 186601 (2013).

\bibitem{Korenivski2013:EPL}
V. Korenivski, A. Iovan, A. Kadigrobov, and R.~I. Shekhter,
Europhys. Lett. {\bf 104}, 27011 (2013).

\bibitem{Naidyuk2012:NJP}
Y. G. Naidyuk, O. P.  Balkashin, V.~V. Fisun, I.~K. Yanson, A. Kadigrobov, R.~I. Shekhter,
M. Jonson, V. Neu, M. Seifert, S. Andersson, and V. Korenivski,
New J. Phys. {\bf 14}, 093021(2012).

\bibitem{Fabian2007:APS}
J. Fabian, A. Mathos-Abiague, C. Ertler, P. Stano, and I. \v{Z}uti\'c,
Acta Phys. Slov. {\bf 57}, 565 (2007).

\bibitem{Hanbicki2002:APL}
A.~T. Hanbicki, B.~T. Jonker, G. Itskos, G. Kioseoglou, and A. Petrou,
Appl. Phys. Lett. {\bf 80}, 1240 (2002).

\bibitem{Zega2006:PRL}
T.~J. Zega, A.~T. Hanbicki, S.~C. Erwin, I. \v{Z}uti\'c, G. Kioseoglou, C.~H. Li, B.~T. Jonker, and R.~M. Stroud,
Phys. Rev. Lett. {\bf 96}, 196101 (2006).

\bibitem{Salis2005:APL}
G. Salis, R. Wang, X. Jiang, R.~M. Shelby, S.~S.~P. Parkin, S.~R. Bank, and J.~S. Harris,
Appl. Phys. Lett. {\bf 87}, 262503 (2005).

\bibitem{Koshihara1997:PRL}
S. Koshihara, A. Oiwa, M. Hirasawa,  S. Katsumoto, Y. Iye, C. Urano, H. Takagi, 
and H. Munekata, %I13 C. Urano not S. Urano
 %Ferromagnetic order induced by photogenerated carriers in magnetic {III-V}
%semiconductor heterostructures of (In,Mn)As/GaSb
Phys. Rev. Lett. {\bf 78}, 4617 (1997).

\bibitem{Ohno2000:N}
H. Ohno,  D. Chiba,  F. Matsukura, T. Omiya, E. Abe,  T. Dietl , Y. Ohno and K.
Ohtani,
 %Electric-field control of ferromagnetism
Nature {\bf 408}, 944 (2000).

\bibitem{Dietl2014:RMP} T.~Dietl and H.~Ohno,
%Dilute ferromagnetic semiconductors: Physics and spintronic structures
Rev. Mod. Phys. {\bf 86}, 187 (2014).

\bibitem{DiasCabral2011:PRB}
E. Dias Cabral, M.~A. Boselli, R. Oszwaldowski, I. \v{Z}uti\'{c}, and
I.~C. da Cunha Lima, Phys. Rev. B {\bf 84}, 085315 (2011); 
L.~D. Anh, P.~N. Hai, Y. Kasahara, Y. Iwasa, and M. Tanaka, {\em ibid.} {bf 92}, 161201(R) (2015).

\bibitem{Wang2006:JPCM}
J. Wang, C. Sun, Y. Hashimoto, J. Kono, G.~A. Khodaparast, 
L.  Cywi\'{n}ski, L.~J. Sham, G.~D. Sanders, C.~J. Stanton, and H. Munekata,
%Ultrafast magneto-optics in ferromagnetic IIIÐV semiconductors,
J. Phys.: Condens. Matter {\bf 18}, R501 (2006).

\bibitem{Lazic2016:PRB}
P. Lazi\'{c}, K.~D. Belashchenko, and I. \v{Z}uti\'{c},
Phys. Rev. B {\bf 93}, 241401(R) (2016).


\end{thebibliography}
\end{document}